\documentclass[prb,twocolumn,showpacs,aps]{revtex4}

\usepackage{graphicx}
\usepackage{dcolumn}
\usepackage{amsmath}
\usepackage{amssymb}

\begin{document}

\title[Short Title]{$^{ 207}$Pb AND $^{17}$O NMR STUDY OF THE ELECTRON DENSITY 
DISTRIBUTION IN METAL PHASE OF BaPb$_{1-x}$Bi$_x$O$_3$}  

\author{Yu.~Piskunov, A.~Gerashenko, A.~Pogudin, A.~Ananyev, K.~Mikhalev,
K.~Okulova}
\affiliation{Institute of Metal Physics,UB RAS,Ekaterinburg,Russia}

\author{S~.Verkhovskii}
\altaffiliation[Corresponding author:]{Dr. S. Verkhovskii,
                                      Institute of Metal Physics, UB RAS,        		                                 Kovalevskaya str., 18, 
                                  		Ekaterinburg 620219, Russia}
\altaffiliation[FAX: ]{+ 7 3432 - 745244}
\email[E-mail: ]{Verkhovskii@imp.uran.ru}
\affiliation{Institute of Metal Physics,UB RAS,Ekaterinburg,Russia}

\author{A.~Yakubovsky}
\affiliation{Russian Research Centre "Kurchatov Institute", Moscow, Russia}

\author{A.Trokiner}
\affiliation{Laboratoire de Physique du Solide,  E.S.P.C.I., Paris, France}

\smallskip
\begin{abstract}
The $^{17}$O and$^{ 207}$Pb NMR spectra were measured in ceramic 
samples in the metallic phase of BaPb$_{1-x}$Bi$_x$O$_3$  oxides
$(0<x\leq 0.33)$. The inhomogeneous magnetic broadening which appears due
to a distribution of the Knight shifts was analyzed in detail. It is shown
that Bi atoms, which are randomly incorporated in BaPbO$_3$ parent compound
give rise to an increased conduction electron spin density within  an area
which is delimited by its two first cation shells. According to NMR data
the percolative overlap of these areas occurs in superconducting
compositions and  it is accompanied by a sharp growth of the average
Knight shift $<K_s>$. The decrease of $<^{17}K_s>$ with temperature
revealed for $x=0.33$ evidences for an opening of the energy gap near 
$E_F$ near the metal-semiconductor transition ($x=0.35$).
\end{abstract}

\pacs{71.30.+h, 74.24.Jb, 76.80.-g, 76.60.-k}

\maketitle
\smallskip
\section{Introduction}

During the last two decades many studies have been devoted to the
electron states in the conduction band of superconducting 
BaPb$_{1-x}$Bi$_x$O$_3$ oxides (BPBO). Substituting Bi for Pb in the
metallic parent compound, BaPbO$_3$, leads to superconducting compounds
in the composition range $0.1<x<0.35$ \cite{1} (superconducting 
composition range, SCR). The density of states at Fermi level $N(E_F)$
is found to be small and $T_c$, the superconducting transition 
temperature, estimated from thermal capacity  data in the Debye 
approximation is substantially smaller than observed $T_c$ \cite{2,3,4}. According to extended photo-emission studies \cite{5,6,7} a pseudo
gap near $E_F$ appears for $x \geq 0.2$ and exists as a real gap
for semiconducting compositions ($x>0.35$). Finally, 
the end member of the family, BaBiO$_3$, is an insulator with an 
energy gap $ \sim 1eV$ due to the commensurate charge density wave 
(CDW) breathing mode ($ \textbf{qa}= \{\pi;\pi ;\pi\}$)  developed in 
the BiO$_6$-octahedra sublattice. It was 
considered  that electron-phonon coupling might be increased 
due to the breathing 
phonon mode existing in SCR as the short-wave charge fluctuations \cite{8}.
Exclusively electronic origin of the attractive retarded coupling
between the carriers due to the 
\textit{"skipped valence"} effects of Bi was proposed by Varma 
\cite{9} and the relevant phase diagram was developed in \cite{10}
 for semiconducting phase of bismuthates. However 
the stability of a uniform electron system in a metallic phase in the presence of 
random ionic potential due to the cation disorder is still an open question. As noted 
in Ref.~\onlinecite{11,12,13,14,15} the changes in the carriers motion near the more electronegative Bi 
cations can lead to an electronic ground state associated to a non uniformly 
distributed electron system in the real space.

Most of the experimental data concern parameters describing the electronic and 
structural properties of an averaged crystal. The static and dynamic effects of the 
cation disorder in the sublattice of Pb(Bi)O$_6$  octahedra on conduction electrons 
are considered only in few publications. In discussing neutron diffraction results it 
was concluded \cite{16} that in the range of $0.12<x<0.3$, the crystal structure of 
BaPb$_{1-x}$Bi$_x$O$_3$  can be considered as a micro-dispersed mixture of 
tetragonal and orthorhombic phases. Furthermore, a reversible change of the relative 
fraction of the two phases occurs with temperature. EXAFS \cite{17} and $^{137}$Ba 
NMR\&NQR \cite{18} studies have confirmed the presence of local lattice distortions in 
the Bi(Pb)O$_6$ sublattice. The influence of short-range ordering of Bi cations in 
regard to the developed electron instability was studied in Ref.~\onlinecite{19}. It was shown that 
some specific configurations of cations give rise to non-metallic cluster for metallic 
compositions $x>0.1$. 

Being a local probe of electronic properties, Nuclear Magnetic Resonance 
(NMR) is well suited to determining the spatial distribution of electric
and magnetic fields of non homogeneous structures. The Knight shift
$^{207}K_s$ and spin-lattice relaxation rate $^{207}T_1^{-1}$ of lead
were measured at $T=4K$ in BaPbO$_3$. The behavior of these properties 
is similar to the one of normal metals \cite{20}. $^{17}$O \cite{21} and 
$^{ 207}$Pb \cite{22} NMR studies show a monotonous increase of the Knight shift
$K_s \sim N(E_F)$ as a function of $x$ when approaching the 
metal-semiconductor transition and no evidence for electron correlation 
effects was reported for composition $x=0.25$ \cite{22} which corresponds to 
the highest $T_c$, $T_{c,max}=12.5K$. It is worth to note that 
aforementioned NMR data were analyzed assuming that all the oxygen 
and lead ions are located in magnetically equivalent sites of the 
ideal crystal, ignoring any disorder. Only an abnormally broad distribution
of the $^{207}$Pb NMR line shift in SCR for $0.10< x<0.18$ \cite{22}
was attributed to the effect of CDW instability developed in the 
oxides placed compositionally close to the 
metal-semiconductor transition boundary ($x =0.35$).

Taking into account these results we carried out a detailed study of $^{17}$O and 
$^{ 207}$Pb NMR for BaPb$_{1-x}$Bi$_x$O$_3$  in the metallic region ($0 \leq 
x \leq 0.1$) as well as in the superconducting region ($0.1\leq  x \leq $0.36). The 
analysis of NMR spectra evidences the formation and evolution of a microscopic 
inhomogeneous distribution of the charge  and spin densities of mobile carriers in 
these compositions.  

\smallskip
\section{Experimental}

The measurements were performed on single-phase powder samples 
BaPb$_{1-x}$Bi$_x$O$_3$  (BPBO) in the range, $0\leq x\leq0.36$). The BPBO 
samples were prepared with the following compositions, 
$x=0.36, 0.33, 0.27, 0.21,0.18, 0.15, 0.12, 0.09, 0.03$  and 0. 
The samples were obtained by the conventional 
solid-state reaction in air similar to that described in \cite{23}. We put a lot of care into 
avoiding any macroscopic heterogeneity due to Pb(Bi) substitution. After a 
preliminary grinding, some of the samples were enriched with the $^{17}$O isotope. 
The powder was poured into a platinum cup and placed into a quartz tube with 
flowing $^{17}O_2 $ mounted in the furnace. During the heat treatment the oxygen 
gas in the quartz loop was continuously scrubbed with Ascarite and cold plate (at 
$180K$). The $^{16}O-^{17}$O isotope substitution was carried out at $650 
^{\circ}C$ for 168 hours under an oxygen pressure of 730 Torr. The system was 
filled twice with "fresh" $^{17}O_2$. The sample was then cooled to $T=300K$ 
with a rate of $15K/hour$. According to mass spectroscopic analysis, the final 
$^{17}$O enrichment of the sample was about 15\%.
During the preparation and the enrichment the cooling rate used was always 
slow enough to obtain samples with the 
minimal amount of oxygen vacancies.

The x-ray powder diffraction measurements were performed at $T=300K$ 
with a DRON4-A diffractometer with CuK$_{\alpha}$ radiation in the range of 
2$\theta$=(10-70) degrees. To minimize the instrumental error in determining the 
positions of Bragg reflexes the sample was mixed with a powder of crystalline 
germanium. The best fit of experimental x-ray diffraction patterns was found when 
assuming orthorhombic symmetry of the unit cell parameters for all compositions. 
No other peaks evidencing the presence of spurious phases in the samples were 
revealed. A monotonous decrease of the unit cell volume is deduced as x increases. 
The structural parameters of the samples are in a good agreement with the data 
published in Ref.\cite{24} for BaPb$_{1-x}$Bi$_x$O$_{3-\delta}$  with a small oxygen 
deficiency, $\delta$. 

\begin{figure}[tp]
\includegraphics[width=0.4\textwidth,viewport=44 81 502 731]{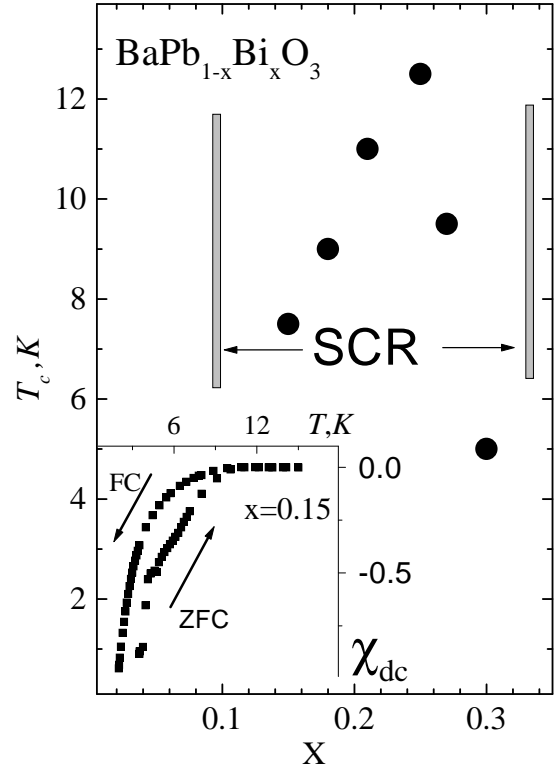}
\caption{$T_c$ as a function of $x$, the bismuth content. The inset shows dc FC and ZFC 
magnetic susceptibility vs temperature plots for sample x=0.85. $\chi_dc$ data are 
normalized to the susceptibility $\chi_{dc}$(2K) of the Pb sample having the very 
same form as the measured pellet of the oxide}
\label{fig1}
\end{figure}

$T_c$ was determined with  ac and dc magnetic susceptibility measurements ( 
Fig~\ref{fig1}). The superconducting transition width does not exceed $2K$ for $x =0.25$. 
The superconducting volume fraction was estimated from the magnetization $M(T)$. 
It was measured by field cooling (FC) and zero field cooling (ZFC) experiments with 
a SQUID-DESIGN device operating at $B_0=10^{-3}T$ (inset in Fig.1). The 
deduced Meissner fraction exceeds 0.8 for $x=0.15$ and $x=0.27$. For $x>0.3$ and 
$x<0.15$ the superconducting response is absent down to $4K$. Concerning the 
homogeneity of composition, the x-ray analysis and dc-magnetization data are in favor of a high enough homogeneity on macroscopic scale for all the samples.

\begin{figure*}[tp]
\includegraphics[width=0.8\textwidth, viewport=20 30 760 560]{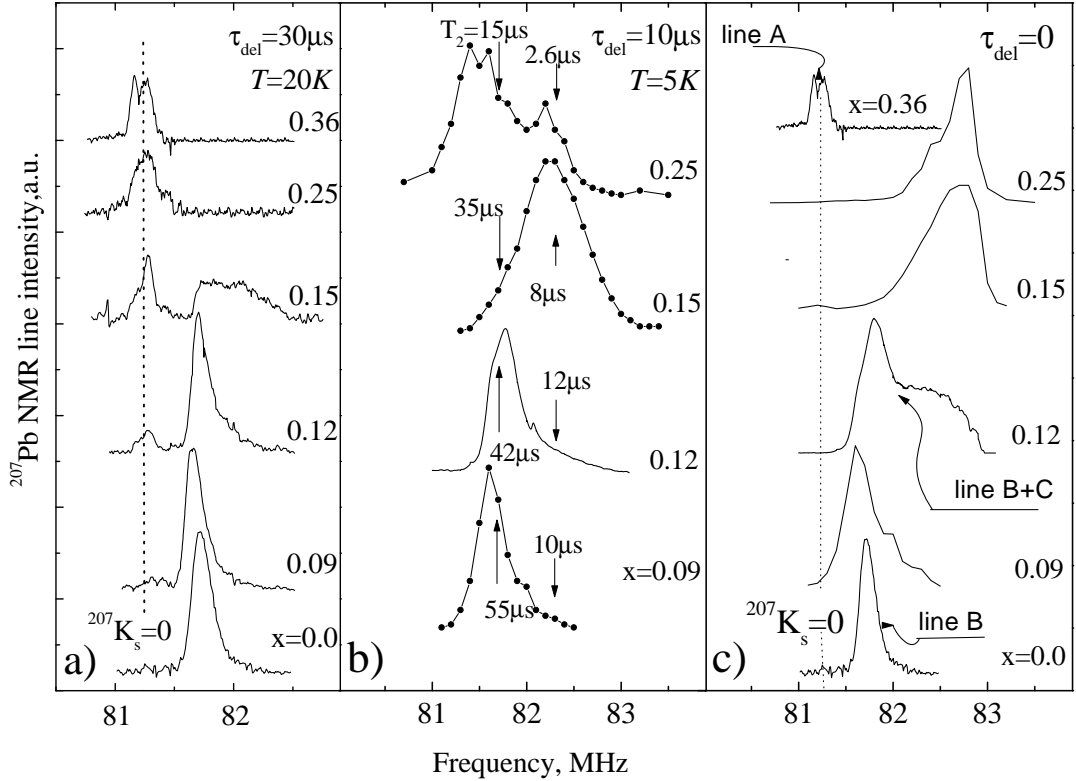}
\caption{$^{ 207}$Pb NMR spectra of BaPb$_x$Bi$_{1-x}$O$_3$ measured by spin echo technique 
with different delay time $\tau_del$ between echo pulses.}
\label{fig2}
\end{figure*}

$^{207}Pb$ and $^{17}$O NMR measurements were carried out on a Bruker NMR 
pulse spectrometers over the temperature range of $10-350K$  in 3 magnetic fields, 
2.0, 9.1 and $11.7T$. The spectra were obtained by Fourier transformation of the 
second half of the echo signal obtained with the $(\pi/2)_x-{\tau}_{del}-
(\pi)_x$ pulse sequence. A phase cycling of the pulses was used in order to 
suppress spurious signals arising from the transient rings in the rf-circuit after the 
pulses. The spectra for which the width exceeds the frequency band excited by the 
rf-pulse were obtained by irradiating the whole spectrum at equidistant frequencies. 
The spectrum was reconstructed after Fourier transformation of the corresponding 
echoes.

The components of the magnetic shift tensor ($K_{iso}$, $K_{ax}$) for 
$^{17}$O and $^{ 207}$Pb as well as the electric field gradient (EFG) parameters (for 
$^{17}$O) - quadrupole frequency, ${\nu}_Q$, and asymmetry parameter, $\eta$, - 
were determined by computer simulation of the measured NMR spectra. The powder 
pattern simulation program takes into account the quadrupole coupling corrections 
up to the second order of the perturbation theory. The NMR powder pattern is 
calculated with a distribution function of the NMR parameters and  a non-uniform 
broadening of the spectra can be specified. 

For $^{17}$O, the positions of the features of both, the central line 
(transition $m=-\frac{1}{2}\leftrightarrow+\frac12$)  and satellite lines 
 ($m=\pm \frac12 \leftrightarrow \pm \frac32$), were analyzed. We used solid 
Pb$^{+II}$(NO$_3$)$_2$ for $^{207}$Pb ($^{207}\nu_L$= 80.978MHz) and H$_2$O 
for $^ {17}$O as frequency references.

Above $T_c$ the magnetic susceptibility  ${\chi}_m$ was measured by a 
Faraday balance technique. The measurements were performed in the temperature 
range of $4.2-370K$ in a magnetic field of $0.5T$ and the instrumental error did not 
exceed 0.02 ${\chi}_m$.

\smallskip
\section{Experimental results and discussion}
\subsection{$^{ 207}$Pb NMR spectra}

\begin{figure}[tp]
\includegraphics[width=0.4\textwidth, viewport=22 80 540 740]{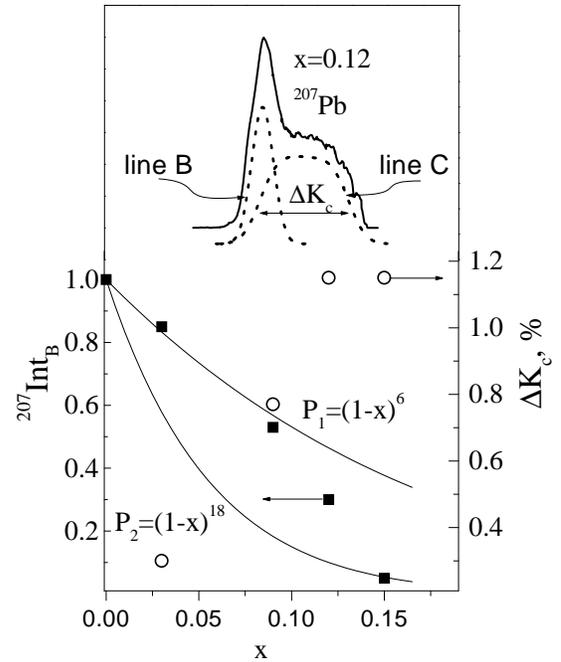}
\caption{ Relative intensity of the narrow peak $^{207}$Int$_B$ and width line C $Delta K_c$  as a function of 
$x$ in $^{ 207}$Pb NMR spectra of BaPb$_{1-x}$Bi$_x$O$_3$ .}
\label{fig3}
\end{figure}

The $^{ 207}$Pb NMR spectra shown in Fig.2 for BPBO samples were 
obtained at $B_0=9.1T$, in the temperature range of $5-20K$. For SCR, spectra 
were measured in the normal state since $B_0$ exceeds the critical field, $H_{c2}$. A 
single non uniformly broadened line (Line A) is found at low frequency in the 
semiconducting region, $x>0.35$, (Fig.2a). Its shift $^{207}K$(0.36)=0.30 (5)\% is in 
the range of chemical shifts reported for Pb$^{+IV}$ in insulating oxides 
Ba$_2$PbO$_4$ (0.21\%) and Sr$_2$PbO$_4$ (0.23\%) \cite{25}. Line A still 
exists in SCR 
but its intensity sharply decreases with decreasing $x$. For $x=0.12$ the relative 
intensity of line A is less than 5\% of the total intensity 
of $^{ 207}$Pb spectrum. 
Even for $x=0.36$, less than one third of the lead atoms contribute to the line A (Fig.2a). Its shift is independent on both, $T$ and $x$. Its spin-lattice relaxation time 
$T_1$ is of order of 1s and the spin-spin relaxation time $T_2$ is about 75(10)$\mu s$ at $T=20K$. At the moment we cannot definitely judge on the origin of this line 
in SCR. The measurements on semiconducting compositions of BPBO ($0.4 \leq x 
\leq 0.9$) are planned to clear up this point.

On the end of the phase diagram, in BaPbO$_3$, a single, nearly gaussian, 
symmetric line (line B) is detected at higher frequency (Fig.2). Its shift, 0.91(2)\% is 
$T$-independent and is in quite a good agreement with published data in \cite{20,21,22}. 
As shown in Fig.2b, the high-frequency part of the NMR spectra in the Bi-containing 
compounds depends strongly on the delay time $ \tau_{del}$ between rf-pulses. In 
order to analyze the spectra obtained in metallic compounds ($x<0.35$), let us focus 
on the main features of the  $^{207}T_2$ behavior in the high frequency part of the 
spectra. 

(i) In BaPbO$_3$ the value of   $^{207}T_2^{-1}$ exceeds greatly  the homonuclear 
magnetic dipole contribution $^{207}T_{2,dip}^{-1}\approx(0.1-0.3)ms^{-1}$ 
estimated for sample with natural abundance of the lead isotopes. In 
BaPbO$_3$ and $x=0.09$ samples,  $^{207}T_2(5K)=55(5) \mu s$. Besides, 
the intensity of the echo, $I(2\tau_{del}$), demonstrates an oscillating 
behavior at low temperature with $\omega_{osc}=4.5(5)\cdot 10^5s^{-1}$. The 
homonuclear indirect coupling term $J{ex}$I$_{z1}$I$_{z2}$  is assumed to be 
responsible for the  $^{207}T_2$ in the metallic phase. The exchange integral for 
BaPbO$_3$ estimated within the Ruderman-Kittel model, $J_{ex}=0.2\cdot10^{-21}erg$, was found to be consistent with the measured value of $\omega_{osc}$.

(ii) As shown on Fig.2b,   $^{207}T_2$ changes within the non-uniformly 
broadened spectra. The arrows point to the operating frequencies. The data can 
be fitted with the following expression  $^{207}T_2(\nu)^{-1}_{exp}\sim 
^{207}K_s^{\alpha}(\nu)$ where $\alpha=1.0-1.3$ for $x<0.2$. The Knight 
shift $^{207}K_s(x)$ for a fixed $x$, was defined as follows. We have assumed 
that the chemical part of the total shift does not change with $x$ since the 
valence state of lead is fixed (Pb$^{+IV}$). Thus, we write :
\begin{eqnarray}
^{207}K_s(x)&=&\{^{207}K(x)-{^{207}}K(x=0.36)\}\nonumber \\
            &=&\frac{1}{\mu_B}^{207}H_{hf}\chi_{s,loc} 
\end{eqnarray}
 $^{207}H_{hf}$ is hyperfine magnetic field  arisen due to hyperfine  interaction  
with conducting electrons, which spin density near the given atom can be 
characterized by local spin susceptibility  $\chi _{s,loc}$.

(iii) The $T_2(\nu )$ values decrease sharply with increasing x on approaching the 
metal-semiconductor  transition.

Taking into account the variation of $T_2$ within a spectrum, we have 
restored the real $^{ 207}$Pb NMR line shape in the following way (Fig.2c): the 
intensity of echo signals measured at each frequency, $\nu$  , with  $\tau_{del}=10\mu s$ was multiplied by the factor $exp(2\tau_{del}/{T_2(\nu)})$. 

Comparing the restored line shapes ( $\tau _{del}=0$) to the spectra measured 
with  $\tau _{del}=10 \mu s$, one may see that line B corresponding to BaPbO$_3$ 
is still present in $x=0.09$ and 0.12 compounds. Nevertheless, the corresponding 
Knight shift shows a small increase as x increases. Line B is located on a low-
frequency edge of a broad line (line C). Line C shifts toward higher frequency as $x$ 
increases. On the SCR boundary, for $x=0.12$, the resulting spectrum has the best 
resolved structure (Fig.2c) since line B and C have more or less the same weight. 
For $x \geq 0.15$, mainly line C is present, line B has almost disappeared. The 
dominant intensity of line C in SCR corresponds to $^{207}K_s>1.5$\%. 

Thus, substituting Bi in BaPbO$_3$ produces a large broadening of the 
spectra due to a distribution of the Knight shifts. The broadening is accompanied by 
a distribution of $^{207}T_2^{-1}(\nu)$ and $^{207}T_1^{-1}(\nu)\sim ^{207}K_{s}^{2}(\nu)$. It reveals a non uniform $(q\neq  0)$ distribution of the carrier 
density "depending" on the local environment of the lead atoms.

Let us focus on the non superconducting metallic compounds, $x < 0.12$. 
Since $x$ is rather small, it is reasonable to assume a random distribution for Bi ions 
in the Pb-sites. As shown on Fig.3, the measured relative intensity of line B 
($^{207}$Int$_B$) decreases while the width of line C ($ \Delta K)_C$ increases with 
increasing $x$. In parallel, for line C, the relative intensity as well as the shift of the 
center of gravity (Fig.2c) increases as the Bi concentration increases. Thus, Pb 
atoms located in an area around Bi experience a high local spin susceptibility  $\chi_{s,loc}$. The characteristic radius of this area can be estimated on the assumption of 
a binomial distribution $P_i(x)$ which valid only for Bi-diluted oxides. On Fig.3, 
the upper curve $P_1(x)=(1-x)^6$ corresponds to the fraction of Pb having no Bi 
atoms in the nearest cation shell. The bottom curve $P_2(x)=(1-x)^{18}$ is the 
fraction of Pb having no Bi atoms in both nearest and the next nearest cation shells. 
The experimental data $^{207}$Int$_B$ are well situated between these two curves, 
$P_1(x)$ and $P_2(x)$, which means that the radius of the area with high  $\chi 
_{s,loc}$  around Bi, line C being attributed to Pb atoms in these areas, does not 
exceed 1-2 lattice constants. A steeper decrease of $^{207}$Int$_B$ occurs for $x \geq 
0.09$. It is interpreted as due to the percolative overlap of the Bi-containing areas 
expected for $x \geq 0.12$ near the transition to SCR

In SCR, the increase of the relative intensity of line C (Fig.2c) evidences that 
most of Pb nuclei are in the sites with high  $\chi _{s,loc}$. 

$^{ 207}$Pb NMR spectra were  available for measurements up to room 
temperature only for $x \leq 0.15$. It was found that the line position, the total width 
at half height ($\Delta K_s)_{0.5}$ and $T_2$ measured at a given $K_s$ are roughly 
independent of $T$. Unfortunately on the other end of the studied phase diagram i.e. 
for $x\leq 0.25$, no echo signal was detected in the range of $^{207}K_s>0.5\%$, 
probably due to very short  $^{207}T_2<1 \mu s$.

\smallskip
\subsection{$^{17}$O NMR spectra}

The oxygen atoms are located at the apex of octahedra  with Pb or Bi at the 
centers. The resonance frequency of $^{17}$O NMR-probe is determined by both the 
hyperfine magnetic interaction and the quadrupolar interaction.

\begin{figure}[!]
\includegraphics[width=0.4\textwidth]{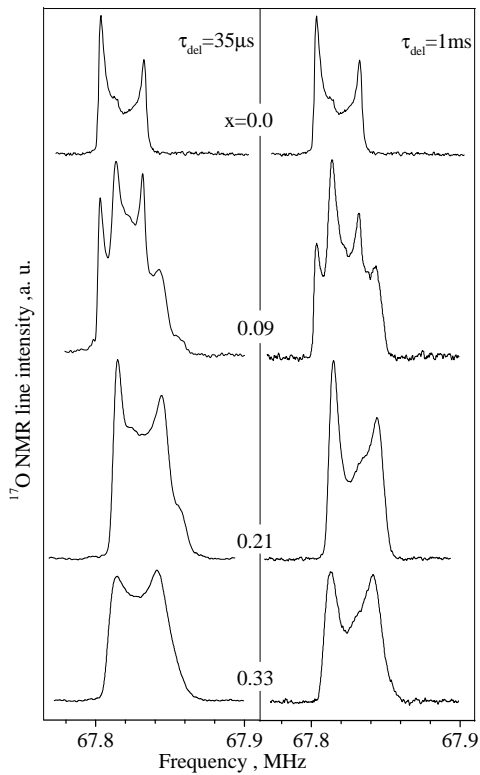}
\caption{ $^{17}$O NMR spectra (transition $m=-\frac12\leftrightarrow \frac12 $) measured at room 
temperature in BaPb$_{1-x}$Bi$_x$O$_3$  by spin echo technique with 
different $\tau _del$.}
\label{fig4}
\end{figure}

The $^{17}$O central lines (transition $m_I=-\frac12 \leftrightarrow 
+\frac12$) measured at room temperature at $B_0=12T$ are presented in Fig.4. 
The satellite lines corresponding to the transitions $m_I=\pm \frac12 
\leftrightarrow \pm \frac32$ were also measured (Fig.6). Simulation of the 
central NMR line shapes measured at different magnetic fields allowed us to 
determine with a reasonable accuracy the components $^{17}K_{iso}$, 
$^{17}K_{ax}$ of the magnetic shift and ${\nu}_Q$, $\eta$ the EFG tensor. 
${\nu}_Q$, $\eta$ are related to the corresponding Cartesian components $V_{ii}$ of 
electric field gradient \cite{26}:
\begin{eqnarray}
    &{\nu}_Q=&\frac{3e^2Q}{2I(2I-1)hV_{ZZ}};
    \eta=\frac{V_{XX}-V_{YY}}{V_{ZZ}}\nonumber \\
     &with&|V_{ZZ}| \geq |V_{XX}|\geq |V_{YY}|     
\end{eqnarray}
For both parent compounds, BaPbO$_3$ and BaBiO$_3$, all the oxygen sites are 
equivalent since the corresponding spectrum can be described with a single set of 
magnetic shift and EFG. The two-peaked line shape is due to the second order 
quadrupole effect on a powder spectrum (see BaPbO$_3$ on Fig.4). The symmetry 
of the EFG tensor is almost axial. It should be noted that the values of $K_{iso}$ 
and ${\nu}_Q$  listed in Table 1 are consistent with the results reported in Ref.~\onlinecite{21}.

Magnetic and charge equivalence of the oxygen sites is destroyed in the 
intermediate metallic compositions for which spectra show a rather complex 
structure (Fig.4), each peak or shoulder having its own dependence on  $\tau_{del}$ 
and on the repetition time. The intensity of the high frequency part of the spectrum 
grows with increasing $x$, correspondingly $(T_2)^{-1}$ is also larger. These changes 
evidence that in the Bi-containing oxides there is a distribution of the fluctuating 
hyperfine magnetic fields responsible for processes of spin-lattice and spin-spin 
relaxation at the oxygen sites.

The less distorted line shapes are obtained for the shortest  $\tau_{del}$ value,
$\tau_{del}=35\mu s<<(^{17}T_2)_{min}\approx 600 \mu s$. The powder spectra of the central transition measured at $B_0=12T$ is 
rather well resolved in the Bi-diluted oxides, $x=0.09$ (Fig.4). This spectrum is the superposition of three lines, the 
deconvolution is shown in Fig.5a. 

\begin{figure*}[!]
\includegraphics[width=0.8\textwidth]{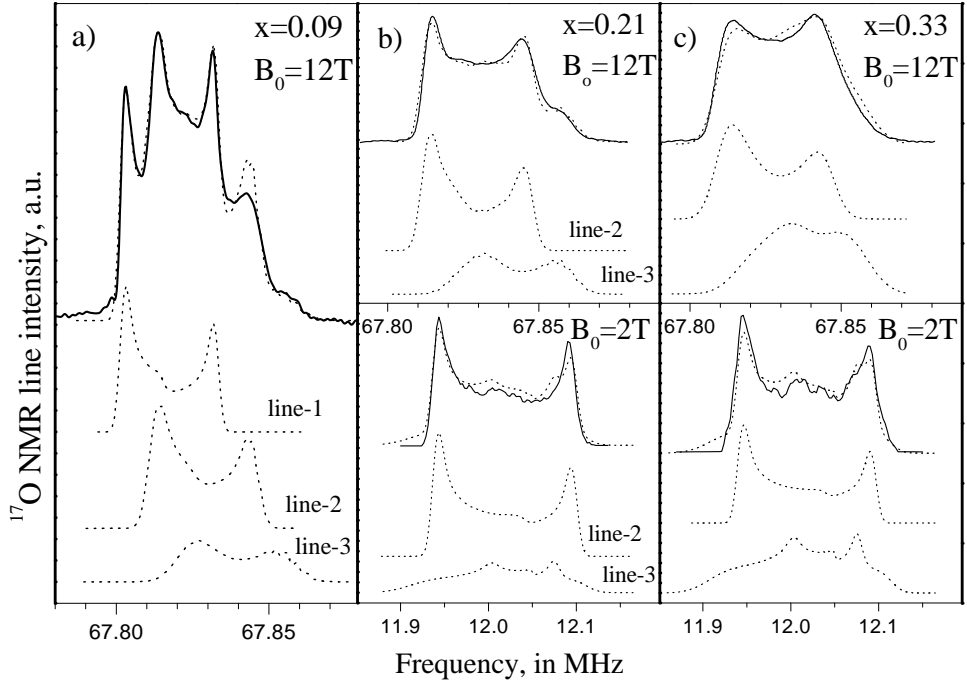}
\caption{ Experimental and simulated (dotted lines) room temperature $^{17}$O NMR 
spectra (transition $-\frac{1}{2}\leftrightarrow \frac{1}{2}$) in low ($B_0$=2T) and high ($B_0$=12T) 
magnetic field.}
\label{fig5}
\end{figure*}

All the simulated lines are broadened by convoluting with a Gaussian function. 
The function is specified in terms of isotropic shift and quadrupole frequency and its 
width at a half of height ($\delta K_{iso}$) or/and ($\delta \nu_Q$). The EFG 
parameters were deduced for each line from the central line measured at low field, 
$B_0=2T$ (Fig.5) and the satellite lines (Fig.6). The calculated spectra are shown by 
dotted lines in Fig.5 and the relevant shift and EFG parameters are listed in Table 1. 
For all the metallic compositions the spectra can be simulated with the same three 
lines, by varying the relative intensities of these lines. Only $\eta$ was adjusted for 
line 3 with the largest magnetic shift. These three lines correspond to three oxygen 
sites differing mainly by their magnetic hyperfine fields.

\begin{figure}[!]
\includegraphics[width=0.5\textwidth,viewport=60 20 760 500]{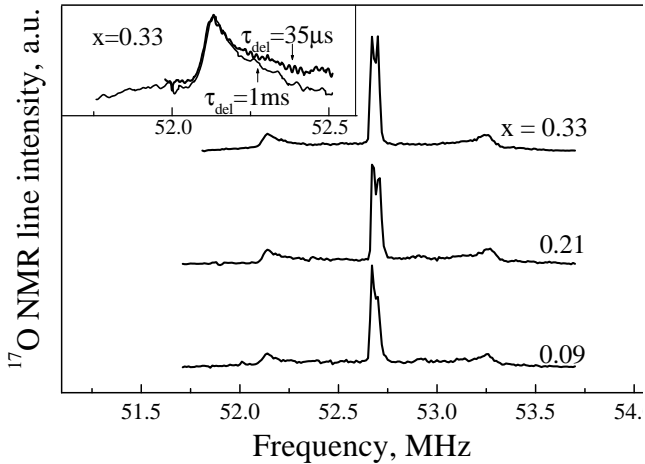}
\caption{$^{17}$O NMR powder patterns for the central line and two first satellites 
measured in BaPb$_{1-x}$Bi$_x$O$_3$  at room temperature.}
\label{fig6}
\end{figure}

In order to understand the origin of the three lines we have analyzed the 
variation of the spectra as a function of $x$. The tensor components of line 1, 
$^{17}K_{iso}$,$^{17}K_{ax}$, ${\nu}_Q$ and $\eta$ are very close to those 
obtained for pure BaPbO$_3$. Its intensity decreases as Bi concentration increases. 
Line 2 and 3 appear only when Bi cations are present. Considering the measured 
relative intensities, $^{17}Int_i(x)$, plotted on Fig.7, it is quite natural to assume that 
line 1 originates from oxygen with no Bi in the nearest and next nearest shells 
whereas line 2 and 3 correspond to oxygen with Bi. Furthermore, the function $(1-
x)^{10}$ (fig.7, dashed curve), which represents the binomial probability of the cation 
configuration without Bi in both, the nearest and the next nearest cation shells 
around oxygen, fits very well $^{17}Int_1(x)$. Thus line 1 is due to oxygen with no 
Bi in its two first shells.

\begin{figure}[!]
\includegraphics[width=0.5\textwidth,viewport=60 50 722 520]{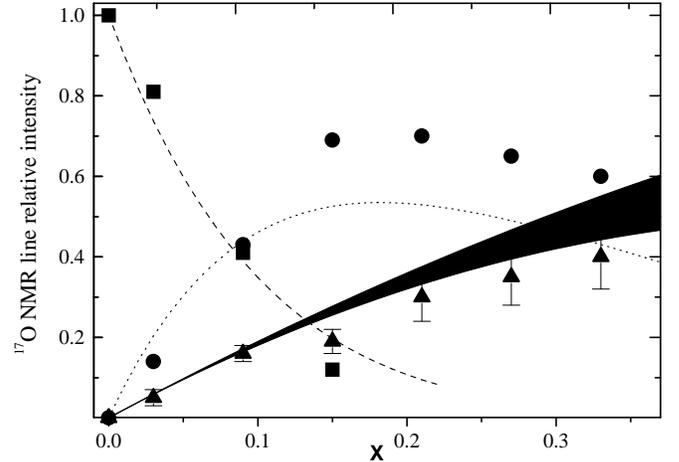}
\caption{Relative intensity of the lines, $^{17}Int_1$, $^{17}Int_2$ and $^{17}Int_3$ of the $^{17}$O NMR 
spectra of BaPb$_{1-x}$Bi$_x$O$_3$  as shown in Fig.5: $\blacksquare$ - line 1, $\bullet$ - line 
2,  $\blacktriangle$ - line 3.}
\label{fig7}
\end{figure}

The shaded region in Fig.7 shows the range of the expected relative intensity 
from the binomial model for different Bi-containing configurations of the first cation 
shell around a given oxygen. The bottom boundary curve, $2x(1-x)$ corresponds to 
the configuration with a single Bi ion in the nearest cation shell whereas the top 
boundary curve, $x(2-x)$, relates to the configuration for which one or two Bi ions 
are present in the nearest cation shell. As seen on Fig.7, for small $x, x<0.1$,  
$^{17}Int_3(x)$ is well described in our model. Thus line 3 with the largest shift can 
be attributed to oxygen with at least one Bi as nearest neighbor. 

The small deviation of $^{17}Int_3(x)$ from the binomial predictions for 
superconducting oxides $(x>0.1)$ is believed to originate from the overlapping of 
the areas with the Bi stimulated high electron density. In the overlapped areas the 
Bi-Bi distance is less than two lattice constants and the foregoing simple binomial 
approach becomes invalid in SCR. It should be added that the reduction of the EFG 
symmetry for line 3 (see $\eta$ on Table 1) is in favor with the suggested assignment.

Line 2 dominates in SCR and its intensity reaches a maximum around 
$x=x(T_{c,max})$ followed by a monotonous decrease in going towards the 
semiconducting compositions. This line might be attributed to oxygen with Bi ions 
only located in the second cation shell. The probability of this cation configuration, 
$(1-x)^{2}(1-(1-x)^{8})$, is represented by the dotted line in Fig.7. As it can be seen 
$^{17}Int_2(x)$ is well described in our model for $x<0.1$.

The increase of the  magnetic shift from line 1 to 3 and the larger $(T_2)^{-1}$ 
of line 3 have to be put in parallel with the increase of the shift of $^{ 207}$Pb NMR 
spectra as the Bi content increases (fig. 2c). As for Pb results, this demonstrates that 
the random substitution of Pb by Bi in BaPbO$_3$ changes the electron density  
within two first cation shells.

\smallskip
\subsection{The average Knight shift of $^{ 207}$Pb and $^{17}$O vs x}

	Pb NMR line is broadened and shifted mainly due to the distribution of the 
Knight shift $^{207}K_s$. The average Knight shift $<^{207}K_s>$ is defined as 
the first 
moment of the inhomogeneously  broadened lines $^{207}g(\nu)$ shown in Fig.2c
\begin{eqnarray}
<^{207}K_s>&=&\frac{<\nu-\nu_0(x=0.36)>}{\nu_0(x=0.36)}\nonumber \\
          &=&\frac{1}{\nu_0}\int(\nu-\nu_0) ^{207}g(\nu)d\nu
\end{eqnarray}
$<^{207}K_s>$ is proportional to the uniform part of the spin susceptibility  $\chi 
_s(q=0)$ \cite{27}
\begin{eqnarray}
\frac{\mu_B<^{207}K_s>}{ ^{207}H_{hf}} =\chi _s (q=0)=\frac{2\mu_B^2 
N(E_F)}{1- JN(E_F)}
\end{eqnarray}
where $N(E_F)$ is the bare density of states (DOS) at the Fermi level. As seen 
on Fig.8, $<^{207}K_s>$ sharply increases at the boundary of SCR. It is rather well 
scaled with the dependence of the electronic thermal capacity coefficient 
$\gamma_{el}\sim N(E_F)(1+\lambda)$ on $x$ \cite{2}. $\gamma_{el}$ probes the 
bare DOS multiplied by the mass enhancement factor $(1+\lambda)$.  The 
electron-phonon coupling constant was estimated in Ref.~\onlinecite{28} and $\lambda$ was 
found to increase in going towards SCR. This leads to a more slander x variation 
of $N(E_F)$ when estimated from the heat capacity data \cite{2} as compared to the 
estimation from the Knight shift data. One may assume that ferromagnetic 
fluctuations in the conduction band increases and the increase of  $\chi _s(q=0)$ 
occurs due to nonvanishing Stoner's enhancement factor $JN(E_F)$.

\begin{figure}[!]
\includegraphics[width=0.5\textwidth,viewport=30 30 660 520]{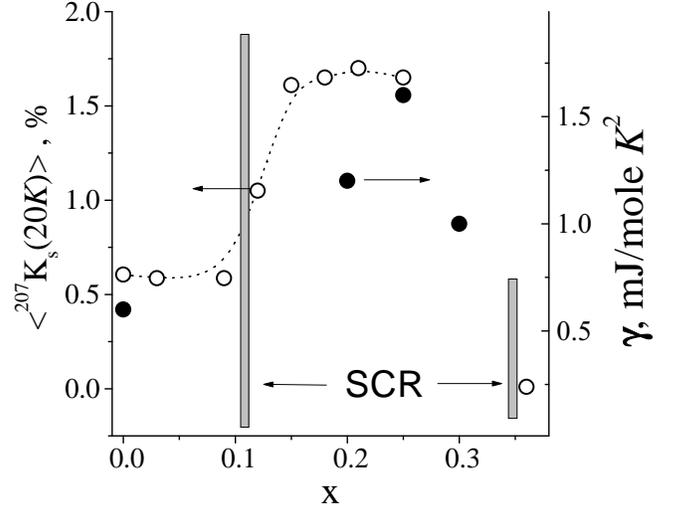}
\caption{ Average Knight shift $<^{207}K_s>$ and the electronic thermal capacity 
coefficient $\gamma _{el}$ \cite{4}	vs x in BaPb$_{1-x}$Bi$_x$O$_3$ .
}
\label{fig8}
\end{figure}

	Let us now consider the average NMR isotropic shift of oxygen 
$<^{17}K_{iso}>$.  It is defined by an expression similar to (4) where $^{17}g(\nu)$ 
is the sum of three Gaussian lines with relative intensity $(^{17}Int)_i$, and 
magnetic shift ($^{17}K_{iso})_i$ corresponding to line 1,2 and 3 respectively. The 
respective widths, ($\delta K_{iso})_i$, of the Gaussian lines  were deduced from 
simulation. The thermal behaviour of $<^{17}K_{iso}>$ down to $T=20$K is 
displayed in Fig.9. For $x < 0.27$, $<^{17}K_{iso}>$ is temperature independent. 
For $x=0.33$, i.e. near the transition to semiconducting state,  a gradual decrease 
of $<^{17}K>$ with decreasing temperature is found. For this sample the static 
magnetic susceptibility is also T-dependent.  In the range of $T=80-300$K the 
measured slope $(\Delta \chi / \Delta T)_{x=0.67}=4.5 \cdot 10^{-11}emu/gr \cdot K$ 
is consistent with the data reported in \cite{29}. Both observations are in favor of the 
opening of a gap near $E_F$ when approaching to the metal-semiconductor 
transition.

\begin{figure}[!]
\includegraphics[width=0.4\textwidth,viewport=25 185 530 700]{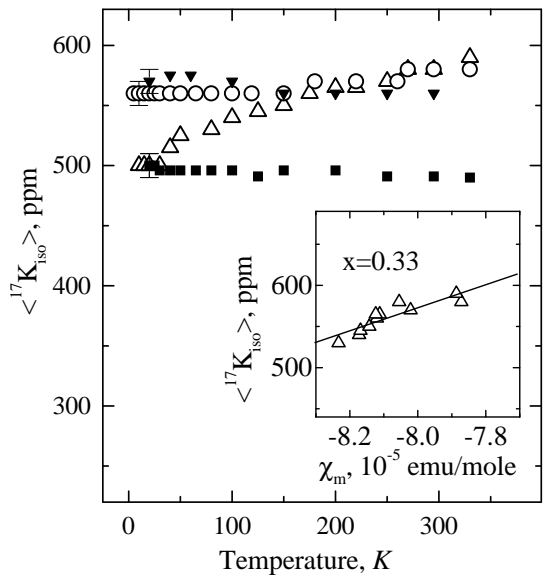}
\caption{Temperature dependence of the average NMR shift of oxygen $<^{17}K>$ in 
BaPb$_{1-x}$Bi$_x$O$_3$: $\blacksquare$~x=0.09,$\blacktriangledown$~-0.21,$\circ$~-0.27, $\vartriangle-0.33$. Inset shows  $<^{17}K>$ vs  magnetic susceptibility  
${\chi}_m$ of the sample x=0.33.
}
\label{fig9}
\end{figure}

	The total magnetic shift  of $^{17}$O NMR line is the sum of  two 
contributions: 
\begin{eqnarray}
<^{17}K>(x,T)=<^{17}\delta>(x) +<^{17}K_s>(x,T)\nonumber \\
             =<^{17}\delta>+\frac{1}{\mu_B}^{17}H_{hf}\chi _s (q=0)	
\end{eqnarray}
where $^{17}K_s$ is the Knight shift and the other shift contributions are 
summarized as "chemical shift" $^{17}\delta$.  For light atoms like oxygen both 
terms in (6) are comparable in magnitude.

For BaPbO$_3$ we have estimated $^{17}K_s(x=0)=130(20)$ppm starting 
from the ratio of $^{ 207}$Pb and $^{17}$O shifts and using eq. (1), (4) and (5). 
The ratio of corresponding hyperfine fields ( $^{207}H_{hf}/^{17}H_{hf})=50(5)$  was 
estimated by taking into consideration the spin-lattice relaxation rate data of 
$^{17}$O and $^{ 207}$Pb \cite{21,22,30}. The deduced  $^{17}\delta$ value is 
$^{17}\delta(x=0) = 220(20)$ppm. Assuming that  $^{207}H_{hf}/^{17}H_{hf}$ is the same 
in all the metallic samples, we have deduced the x dependence of $<^{17}K_s>$ and 
$<^{17}\delta>$ using $<^{207}K_s (x)>$ data up to x=0.25 ( see Table 1). We find that 
$<^{17}\delta>$ is constant within the error bar. 

As shown in the previous section the oxygen with no Bi as the nearest 
neighbor give the main contribution to the $^{17}$O NMR line in superconducting 
samples. Thus one may conclude that no additional paramagnetic (other than 
$\chi _{Pauli}$) contribution to $^{17}\delta$(line-2) appears at these oxygen sites. 

At low temperatures $<^{17}K_s>$ passes through a maximum and then 
decreases for $x=0.33$.  Assuming that for $x=0.33$  $<^{17}\delta>$=200 ppm, the 
average Knight shift is about $<^{17}K_s(x=0.33)>=300$ppm and the 
corresponding bare DOS as 
$N(E_F)_{nmr}=\frac{<^{17}K_s>}{2\mu_B^{17}H_{hf}}=0.16(eV\cdot spin)^{-1}$. It should be noted that 
$N(E_F)_{NMR}$ exceeds more than twice $N(E_F)_{\gamma}$ evaluated from 
the electronic thermal capacity coefficient $\gamma_{el}$ in Ref.~\onlinecite{29} for $x=0.30$, 
which chemical composition  is nearly the same as in BaPb$_{0.67}Bi_{0.33}0_3$. 

\smallskip
\section{Summary}
 
	The NMR spectra of $^{ 207}$Pb and $^{17}$O were measured as a function 
of temperature and x in the metallic phase and in the superconducting BaPb$_{1-
x}$Bi$_x$O$_3$  $(0\leq x\leq0.33)$. Magnetic shift and broadening of the spectra 
were analyzed. It was shown that powder patterns of the spectra are modified due 
to distribution of the Knight shift. The deconvolution of inhomogeneously 
broadened spectra by 3 lines has allowed to reveal a systematic evolution of the 
separate lines intensities with the Bi concentration. Each of these lines was 
attributed to NMR probe with a certain configuration of the neighboring and the 
next neighboring  Pb(Bi) cation shells. 

It was established that in the Bi-diluted oxides the areas with increased 
density of mobile carriers arise around the Bi atoms. The characteristic size of these 
areas does not exceed two lattice constants.  The percolative overlapping of these 
areas is expected at higher Bi concentrations in superconducting BPBO 
compositions.

The temperature dependence of magnetic shift for $^{17}$O NMR lines gives an 
evidence of the gap at $E_F$ in the metal BPBO oxides near the metal to  
semiconductor transition.

\begin{acknowledgments}
	We are very grateful to Dr.A.Inyushkin for $^{17}$O isotopic substitution of the 
samples. We are also indebted to Dr.A.Korolev for advising us about the SQUID-DESIGN device and providing M(T) measurements. This work is supported by 
Russian Found for Basic Researches  (project No 99-02-16974).
\end{acknowledgments}

\begin{table*}[tp]
\caption{$^{17}$O NMR line parameters (for detail see text and Fig.5)}
\begin{center}
\begin{tabular}{|c|c|c|c|c|c|c|c|c|}\hline\hline
  \multicolumn{9}{|c|}{BaPb$_{1-x}$Bi$_x$O$_3$} \\ \hline     
  \makebox[30mm]&\makebox[20mm]{$x$} &\makebox[15mm]{0.0}&
  \makebox[15mm]{0.03}&  \makebox[15mm]{0.09}&  \makebox[15mm]{0.15}&
  \makebox[15mm]{0.21}&  \makebox[15mm]{0.27}&  \makebox[15mm]{0.33}\\   \hline
  &line 1&\multicolumn{4}{|c|}{$\frac{350(10)}{50(10)}$}&\multicolumn{3}{|c|}{--} \\ \cline{2-9}     
  $\frac{^{17}K_{iso}}{\delta K_{iso}},ppm$
  &line 2&--&\multicolumn{6}{|c|}{$\frac{510(15)}{120(20)}$}\\ \cline{2-9}     
  &line 3&--&\multicolumn{6}{|c|}{$\frac{700(20)}{250(50)}$} \\ \hline \hline
  &line 1&\multicolumn{4}{|c|}{-40(10)}&\multicolumn{3}{|c|}{--} \\ \cline{2-9}     
  $^{17}K_{ax},ppm$&line 2 &--&\multicolumn{6}{|c|}{-50(10)} \\ \cline{2-9}
  &line 3&--&\multicolumn{6}{|c|}{-70(10)} \\ \hline  \hline
  &line 1&\multicolumn{4}{|c|}{$\frac{1.13(3)}{0.05(3)}$}&\multicolumn{3}{|c|}{--} \\ \cline{2-9}     
  $\frac{\nu_Q}{\delta \nu_Q},MHz$&line 2 &--&\multicolumn{6}{|c|}{$\frac{1.15(5)}{0.05(3)}$} \\ \cline{2-9}
  &line 3&--&\multicolumn{6}{|c|}{$\frac{1.1(1)}{0.10(5)}$} \\ \hline \hline
  &line 1&\multicolumn{4}{|c|}{$<0.05$}&\multicolumn{3}{|c|}{--} \\ \cline{2-9}     
  $\eta$&line 2 &--&\multicolumn{6}{|c|}{$<0.05$} \\ \cline{2-9}
  &line 3&--& 0.05& 0.015&0.2&0.3&0.35&0.5 \\ \hline \hline
  Relative&line 1&1.0&0.81&0.41&0.12&\multicolumn{3}{|c|}{--}\\ \cline{2-9}
  intensity&line 2 &0&0.14&0.43&0.69&0.7&0.65&0.6\\ \cline{2-9}
  &line 3&0&0.05&0.16&0.19&0.3&0.35&0.4 \\ \hline \hline
  $<^{17}K_{iso}>,ppm$ &300K&350(10)&390(10) &480(20)&520(20)&570(20)&580(20)&590(20)\\ \cline{2-9}
  &20K &350(10)&--&500(20)&--&560(20)&560(20)&500(20)\\ \hline \hline
\end{tabular}
\end{center}
\end{table*}

\end{document}